\documentclass[twocolumn,prb,showpacs,preprintnumbers,amsmath,amssymb,superscriptaddress]{revtex4}
\usepackage{graphicx}
\usepackage{setspace}

\usepackage{color}
\usepackage{ulem} 

\usepackage{amsmath,amsthm,amssymb}
\usepackage{mathrsfs}

\makeatletter
\def\Left#1#2\Right{\begingroup%
   \def\ts@r{\nulldelimiterspace=0pt \mathsurround=0pt}%
   \let\@hat=#1%
   \def\sht@im{#2}%
   \def\@t{{\mathchoice{\def\@fen{\displaystyle}\k@fel}%
          {\def\@fen{\textstyle}\k@fel}%
          {\def\@fen{\scriptstyle}\k@fel}%
          {\def\@fen{\scriptscriptstyle}\k@fel}}}%
   \def\g@rin{\ts@r\left\@hat\vphantom{\sht@im}\right.}%
   \def\k@fel{\setbox0=\hbox{$\@fen\g@rin$}\hbox{%
      $\@fen \kern.3875\wd0 \copy0 \kern-.3875\wd0%
      \llap{\copy0}\kern.3875\wd0$}}%
      \def\pt@h{\mathopen\@t}\pt@h\sht@im%
      \Right}%
\def\Right#1{\let\@hat=#1%
   \def\st@m{\mathclose\@t}%
   \st@m\endgroup}
\makeatother

\usepackage[driverfallback=dvipdfmx]{hyperref}

\begin{document}

\title{Multiple 
amplitude modes in strongly-coupled phonon-mediated superconductors}
\author{Yuta Murakami}
\affiliation{Department of Physics, University of Tokyo, Hongo, Tokyo 113-0033, Japan}
\author{Philipp Werner}
\affiliation{Department of Physics, University of Fribourg, 1700 Fribourg, Switzerland}
\author{Naoto Tsuji}
\affiliation{RIKEN Center for Emergent Matter Science (CEMS), Wako 351-0198, Japan}
\author{Hideo Aoki}
\affiliation{Department of Physics, University of Tokyo, Hongo, Tokyo 113-0033, Japan}
\date{\today}

\begin{abstract}
We study collectively
amplitude modes of the superconducting order parameter in strongly-coupled
electron-phonon systems described by  
the Holstein model 
using 
the non-equilibrium dynamical mean-field theory 
with the self-consistent Migdal approximation as an impurity solver. 
The frequency of the Higgs amplitude mode 
is found to coincide with the superconducting gap 
even in the strongly-coupled 
(beyond BCS) regime. 
Besides the Higgs mode, we find another
collective mode involving the 
dynamics of both the phonons and the superconducting order parameter.
The frequency of this mode, higher than twice the renormalized phonon frequency in the superconducting phase, is shown to reflect a strong electron-mediated phonon-phonon interaction. 
Both types of collective excitations 
contribute to time-resolved photoemission spectra after a strong laser pump as vertex corrections to produce resonance peaks, which allows one to distinguish them from quasiparticle excitations.
\end{abstract}

\maketitle
\section{INTRODUCTION}
Theoretical and experimental investigations of coherent dynamics in superconductors out of equilibrium have a long history.\cite{Anderson1958,Volkov1974,Kulik1981,Sooryakumar1980, Littlewood1981,Littlewood1982,Barankov2004,Yuzbashyan2006a,Barankov2006,Yuzbashyan2006,Papenkort2007,Papenkort2008,Schnyder2011,Carbone2013,Matsunaga2013,Matsunaga2014,Krull2014,Measson2014,Benfatto2014,Kemper2015,Sentef2015,Tsuji2015,Benfatto2015,Capone2015}
Renewed interests have been aroused by  
recent observations of the amplitude mode in conventional phonon-mediated superconductors 
driven by a strong THz laser.\cite{Matsunaga2013,Matsunaga2014}
When a continuous symmetry is broken, there emerge phase modes and amplitude modes.
In a superconductor (SC), where the carriers are subject to the long-range Coulomb interaction, the gapless phase mode couples with the interaction and is lifted to the plasma frequency, 
which is generally much higher than the SC energy scale. 
This mechanism of gaining mass through a coupling with a gauge boson is the so-called Anderson-Higgs mechanism,\cite{Higgs1964,Anderson1963} 
and the unaffected amplitude mode is called the Higgs amplitude mode in analogy with particle physics.
 Before this experiment,\cite{Matsunaga2013} the amplitude Higgs mode had been observed only in a special case, $2H$-${\rm NbSe}_2$, where SC coexists with a charge density wave.\cite{Littlewood1981,Littlewood1982,Sooryakumar1980,Measson2014,Benfatto2014}
Theoretical studies of 
the SC order parameter dynamics 
have so far primarily focused on the static mean-field dynamics.\cite{Anderson1958,Volkov1974,Kulik1981,Littlewood1981,Littlewood1982,Barankov2004,Yuzbashyan2006a,Barankov2006,Yuzbashyan2006,Papenkort2007,Papenkort2008,Schnyder2011,Krull2014,Tsuji2015,Benfatto2014,Benfatto2015,Capone2015}
One important conclusion of 
these works 
is that the frequency of the Higgs amplitude mode ($\omega_{\rm H}$) 
should coincide with the SC gap ($2\Delta_{\rm SC}$) in the BCS regime, \cite{Nambu1985} which is a 
threshold for quasiparticle excitations. 
This relation leads to a suppression of the relaxation channel to Bogoliubov particles and a power-law decay of the Higgs oscillation.

The material used in the recent experiments,\cite{Matsunaga2013,Matsunaga2014} ${\rm NbN}$, has a relatively large dimensionless electron-phonon coupling $\lambda_{\rm eff}\gtrsim1$, corresponding to the strong-coupling regime.\cite{Wolf1985,Dresselhaus1990,Raychaudhuri2008,Demsar2011} Hence it is 
necessary to understand how strong electron-phonon (el-ph) couplings can affect collective excitations in conventional superconductors. 
An important issue 
is the relation between $\omega_{\rm H}$ and the SC gap in the strongly-coupled regime, which  
directly affects the lifetime of the amplitude mode, and therefore its accessibility in experiments.   
In a broader context it is also important to understand effects of the phonon dynamics on the amplitude mode and what type of collective excitations can exist in strongly-coupled el-ph systems.  

Previous studies on collective modes in strongly-coupled phonon-mediated SCs
are limited to very recent works without non-equilibrium phonon dynamics. \cite{Kemper2015,Sentef2015}
In principle, the collective amplitude modes are represented by poles of the dynamical pair susceptibility.
This quantity can be obtained in the strongly-coupled regime by solving the Bethe-Salpeter equation with a frequency-dependent irreducible vertex
on the Matsubara axis and by subsequent numerical analytic continuation for real-frequency information, which would be a bottleneck to this approach.
In this paper, instead of directly solving the Bethe-Salpeter equation, we explore the behavior of the collective modes in strongly-coupled SCs
by simulating the non-equilibrium response to weak perturbations using
the non-equilibrium dynamical mean-field theory (DMFT).\cite{Aoki2013}

\section{MODEL AND METHOD}
The model for strongly coupled SCs that we consider here is the Holstein model,
whose Hamiltonian is
\begin{align}
\mathcal{H}
=&-v\sum_{\langle  i,j\rangle,\sigma}(c_{i,\sigma}^{\dagger}c_{j,\sigma}+{\rm {\rm h.c.}})-\mu\sum_i n_i\nonumber\\
&+\omega_0\sum_i a^{\dagger}_i a_i+g\sum_i (a_i^{\dagger}+a_i)(n_i-1),\label{eq:Holstein}
\end{align}
where $c_i^\dagger$ creates an electron with spin $\sigma$ at site $i$, 
$v$ is the electron hopping, $\mu$ is the 
electron chemical potential, $n_i = 
c_{i,\uparrow}^\dagger c_{i,\uparrow}+
c_{i,\downarrow}^\dagger c_{i,\downarrow}$, 
$\omega_0$ is the bare phonon frequency, $a_i^\dagger$ creates a phonon, 
and $g$ is the el-ph coupling. 

 In DMFT, the lattice model Eq.~(\ref{eq:Holstein}) is mapped onto a single-site impurity model, whose action in the Nambu formalism reads 
  \begin{align}
S_{\rm imp}&=i\int_\mathcal{C} dt dt'  \Psi^{\dagger}(t)\; \hat{\mathcal{G}}^{-1}_{0} (t,t') \;\Psi(t')\nonumber\\
&+i\int_\mathcal{C} dt dt' a^\dagger(t)(i\partial_t-\omega_0)  a(t)\label{eq:imp_action}\\
&-i\int_\mathcal{C} dt g[a(t)+a^\dagger(t)]\Psi^{\dagger}(t) \hat \sigma_3 \Psi(t),\nonumber
\end{align}
where $\int_\mathcal{C}$ denotes an integral on the Kadanoff-Baym (KB) contour, $\Psi^\dagger(t)\equiv[c_{\uparrow}^\dagger(t),c_{\downarrow}(t)]$ a Nambu spinor, $\hat{O}$ a $2\times2$ matrix, and $\hat{\sigma}_{\alpha}$ a Pauli matrix.
 $\hat{\mathcal{G}}^{-1}_{0} (t,t')$ is the Weiss Green's function on the KB contour, which is determined self-consistently so that 
the impurity Green's function, $\hat{G}_\text{imp}(t,t')=-i\langle T_{\mathcal C} \Psi(t) \Psi^\dagger(t') \rangle$, and the impurity self-energy, $\hat{\Sigma}$, coincide with 
the local Green's function for electrons, $\hat{G}(t,t')=-i\langle T_{\mathcal C} \Psi_i(t) \Psi_i^\dagger(t') \rangle$, and the momentum-independent self-energy in the original lattice problem.\cite{Aoki2013}
Here $T_{\mathcal C}$ is the contour ordering operator.
DMFT is justified in the limit of infinite spatial dimensions.  For el-ph systems, we introduce the phonon Green's function defined as $D_\text{imp}(t,t')=-2i\langle T_{\mathcal C} X(t)X(t')\rangle$ with $X=(a^{\dagger}+a)/\sqrt{2}$, and it is equivalent to $D(t,t')=-2i\langle T_{\mathcal C} X_i(t)X_i(t')\rangle$ in the lattice problem.

The most important part in DMFT is how to solve the effective impurity problem.
In principle, even in a non-equilibrium setup, one can solve the problem with a quantum Monte Carlo (QMC) impurity solver.\cite{Aoki2013,Werner2009}
However, because of a dynamical sign problem \cite{Werner2009}
it is difficult to access timescales needed to study the relatively slow dynamics of phonons and order parameters.
In order to avoid this difficulty, we employ the self-consistent (renormalized) Migdal approximation,
\cite{Murakami2015,Bauer2011,Freericks1994,Hewson2002,Capone2003,Hewson2004,Hague2008,Leeuwen2015,Pavlyukh2016}
which is justified when the phonon frequency $\omega_0$ is small compared to the electronic bandwidth.
\cite{Murakami2015,Bauer2011,Freericks1994,Capone2003,Hague2008}
In the self-consistent Migdal approximation 
the electron self-energy ($\hat{\Sigma}$) 
and phonon self-energy ($\Pi$) in the effective impurity model are 
given by
\begin{subequations}\label{eq:Migdal}
\begin{align}
\hat{\Sigma}(t,t')&=ig^2D_{\rm imp}(t,t')\hat{\sigma}_3\hat{G}_{\rm imp}(t,t')\hat{\sigma}_3,\\
\Pi(t,t')&=-ig^2{\rm tr}[\hat{\sigma}_3 \hat{G}_{\rm imp}(t,t')\hat{\sigma}_3\hat{G}_{\rm imp}(t',t)].
\end{align}
\end{subequations}
In equilibrium, we choose the s-wave SC order parameter $\phi\equiv \langle c_\uparrow c_\downarrow\rangle\in {\mathbb R}$.
In the following, we consider an infinitely coordinated Bethe lattice, which has a semi-elliptic density of states, $N(\epsilon)=\frac{1}{2\pi v^2_{\ast}} \sqrt{4v^2_\ast-\epsilon^2}$,
and we set $v_\ast=1$, 
i.e. the electron bandwidth $W$ is 4. We focus on 
half-filling, a small enough phonon frequency $\omega_0=0.4$, and the 
coupling 
regime $\lambda_{\rm eff}\alt2$, where the Migdal approximation should give qualitatively correct results. 
Here $\lambda_{\rm eff}$ is the dimensionless electron-phonon coupling defined from the dressed phonon propagator, see Appendix~\ref{sec:phonon}.
We have confirmed that the results for a lower frequency $\omega_0$ = 0.2 are qualitatively similar to those for $\omega_0=0.4$ in Appendix~\ref{sec:0.2}.

In this paper, we consider two types of excitation protocols. The first protocol is a perturbation 
Hamiltonian $H_{\rm ex}(t)=F_{\rm ex}(t)B_{\bf 0}$ with $B_{\bf 0}=\sum_i(c_{i\uparrow}^\dagger c_{i\downarrow}^\dagger+c_{i\downarrow}c_{i\uparrow})$ and 
$F_{\rm ex}(t)=d_{\rm f}\delta(t)$.  Explanation about the implementation are provided in Appendix~\ref{sec:implement}.
This external field is used to evaluate the dynamical pair susceptibility,
\begin{align}
 \chi^R_{\rm pair}(t-t')=-i\theta(t-t') \langle [B_{\bf 0}(t),B_{\bf 0}(t')] \rangle.
 \end{align} 
We note that this susceptibility is relevant to the dynamics of the amplitude of the SC order parameter, since we take $\phi$ to be real.
\footnote{At half-filling the phase and amplitude of $\phi$ do not mix and $\phi$ remains real even in the non-equilibrium dynamics.}
 In order to obtain the susceptibility we choose a small enough $d_{\rm f}$. 
The second protocol is a modulation of the hopping parameter, $H_{\rm ex}(t)=-\delta v(t) \sum_{\langle  i,j\rangle,\sigma}(c_{i,\sigma}^{\dagger}c_{j,\sigma}+{\rm {\rm h.c.}})$, which mimics the effective band renormalization of a strong and high frequency laser. \cite{Tsuji2011}
   \begin{figure}[t]  
     \centering
     \vspace{-0.cm}
  \includegraphics[width=85mm]{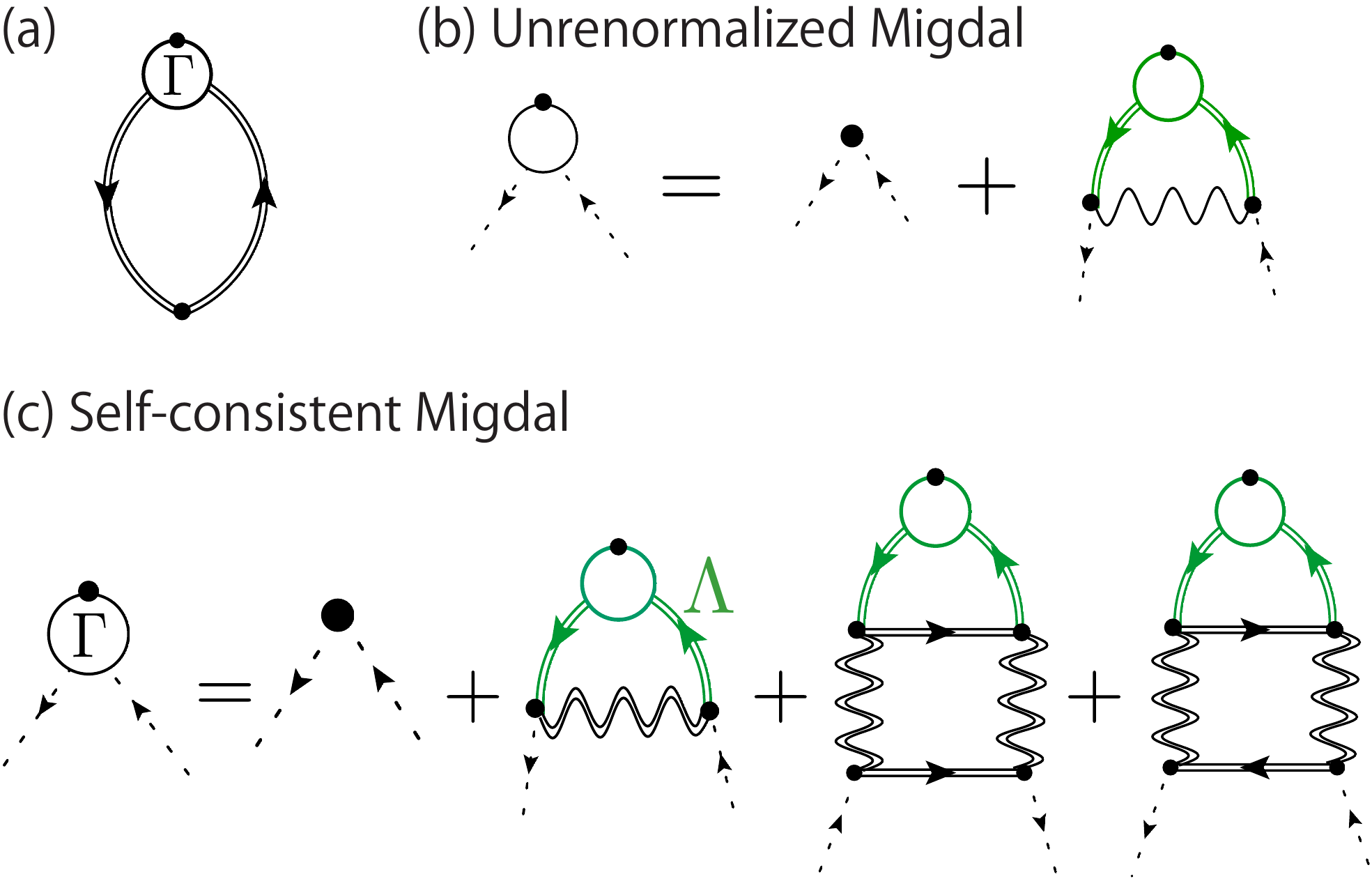}
  \caption{Diagrammatic expressions in the Nambu formalism for (a) the dynamical pair susceptibility, (b) the vertex within the unrenormalized Migdal approximation, and (c) the vertex within the self-consistent Migdal approximation. 
Open circles represent $\hat{\Gamma}$, solid dots $\hat{\sigma}_1$ (bare vertex), and green parts $\hat{\Lambda}$ defined in the text. 
  Solid double lines indicate dressed electron Green's functions, wavy double lines dressed phonon Green's functions, and wavy single lines bare phonon Green's functions.}
  \label{fig:vertex}
  \end{figure}

We now elaborate on the dynamical pair susceptibility evaluated from DMFT+self-consistent Migdal approximation.
In general, we can express the dynamical pair susceptibility as the retarded part of a response function on the KB contour,
\begin{align}
&\chi_{\rm pair}(t,t')\equiv\left.\frac{\delta_\mathcal{C} [-i{\rm tr}\{\hat{\sigma}_1\hat{G}(t,t+0^+_\mathcal{C})\}]}{\delta_\mathcal{C}[F_{\rm ex}(t')]}\right|_{F_{\rm ex}=0}=\nonumber\\
&-i\int_{\mathcal C} d{t_1}d{t_2}{\rm tr} \left[\hat{\sigma}_1 \frac{1}{N}\sum_{\bf k}\hat{G}_{\bf k}(t,t_1)\hat{\Gamma}(t_1,t_2;t')\hat{G}_{\bf k}(t_2,t+0^+_\mathcal{C})\right], \label{eq:pair}
\end{align}
where  ${\bf k}$ is a momentum and $\delta_\mathcal{C}[...]/\delta_\mathcal{C} [...]$ is the functional derivative on the KB contour. 
The diagrammatic expression for $\chi_{\rm pair}(t,t')$ is shown in Fig.~\ref{fig:vertex} (a).
Here $\hat{\Gamma}$ is a renormalized vertex, which can be expressed as 
\begin{align}
\hat{\Gamma}(t,t';t'')=\hat{\Gamma}^{(0)}(t,t';t'')+\left.\frac{\delta_\mathcal{C}[\hat{\Sigma}(t,t')]}{\delta_\mathcal{C} [F_{\rm ex}(t'')]}\right|_{F_{\rm ex}=0},\label{eq:vetex_general}
\end{align}
where $\hat{\Gamma}^{(0)}(t,t';t'')\equiv\hat{\sigma}_1\delta_\mathcal{C}(t'',t)\delta_\mathcal{C}(t'',t')$ is the bare vertex, $\delta_\mathcal{C}(t,t')$ is the delta function on the KB contour, while the second term is the vertex correction.
In perturbative approximations the expression for the self-energy is known, hence we can evaluate $\delta[\hat{\Sigma}(t,t')]/\delta [F_{\rm ex}(t'')]$ explicitly.

In the case of DMFT+self-consistent Migdal approximation, the vertex part is 
given as
\begin{align}
& \hat{\Gamma}(t,t';t'')=\hat{\sigma}_1\delta_\mathcal{C}(t'',t)\delta_\mathcal{C}(t'',t')+ig^2D(t,t')\hat{\Lambda}(t,t';t'')\nonumber\\
&+g^4\hat{\sigma_3}\hat{G}(t,t')\hat{\sigma_3}\int_\mathcal{C}dt_3dt_4 D(t,t_3)D(t_4,t')\times\nonumber\\
&\big\{{\rm tr}[\hat{\Lambda}(t_3,t_4;t'')\hat{G}(t_4,t_3)]+{\rm tr}[\hat{G}(t_3,t_4)\hat{\Lambda}(t_4,t_3;t'')]\big\}.\label{eq:vertex}
\end{align}
Here $N$ is the number of sites, and $\hat{\Lambda}(t,t';t'')\equiv\frac{1}{N}\sum_{\bf k}\int_{\mathcal C} d{t_1}d{t_2}\hat{\sigma_3}\hat{G}_{\bf k}(t,t_1)\hat{\Gamma}(t_1,t_2;t'')\hat{G}_{\bf k}(t_2,t')\hat{\sigma_3}$.
The diagrams for the vertex part $\Gamma$ are displayed in Fig.~\ref{fig:vertex}(c) and a detailed derivation of the expression is given in Appendix~\ref{sec:chi}.

   \begin{figure*}[t]  
     \centering
     \vspace{-0.2cm}
  \includegraphics[width=140mm]{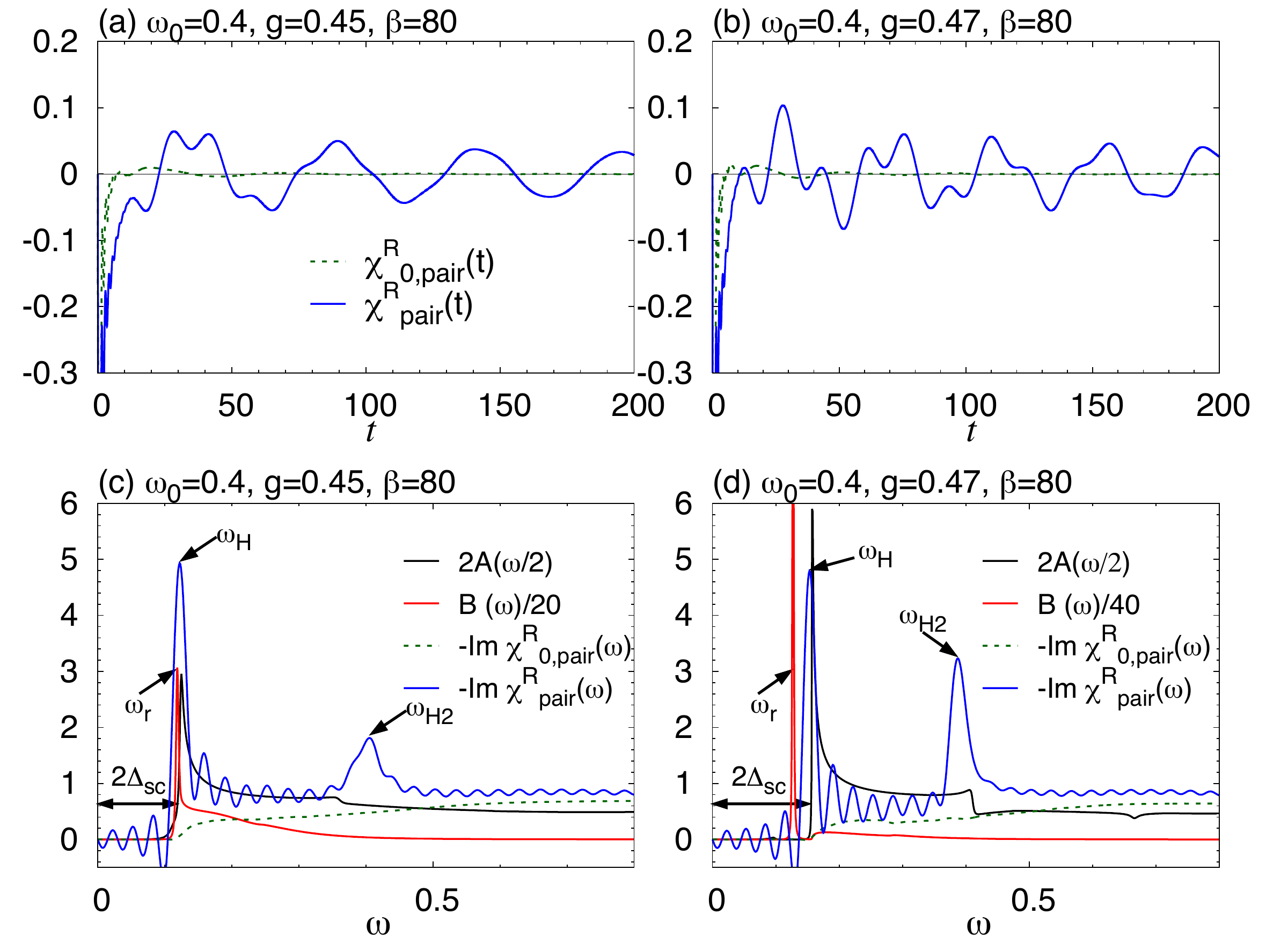}
  \caption{(a)(b) Dynamical pair susceptibility against $t$ evaluated with the full dynamics in the self-consistent Migdal approximation [$\chi_{\rm pair}(t)$] and with 
the bubble diagrams [$\chi_{0,{\rm pair}}(t)$] for $g=0.45, \beta=80$ ($\lambda_{\rm eff}=1.38$) (a) and $g=0.47, \beta=80$ ($\lambda_{\rm eff}=1.89$)  (b). (c)(d) Comparison of
the electron spectrum 
$A(\omega/2)$, the phonon spectrum $B(\omega)$, $-{\rm Im}\chi^R_{0,{\rm pair}}(\omega)$ and $-{\rm Im}\chi^R_{\rm pair}(\omega)$  
 for  $g=0.45, \beta=80$ (c) and $g=0.47, \beta=80$ (d).  $\chi_{\rm pair}(\omega)$ and $\chi_{0,{\rm pair}}(\omega)$ are evaluated from the data at $t\in[0,200]$. The factor of 2 in $A(\omega/2)$ facilitates a comparison between $2\Delta_{\rm SC}$ and $\omega_{\rm H}$. }
  \label{fig:suscep_pair}
  \end{figure*}

In contrast to our treatment, the BCS and unrenormalized Migdal approximations \cite{Kemper2015,Sentef2015} describe a situation where the phonons always stay in equilibrium. 
In these two cases 
the equation for the vertex part contains only the 1st and 2nd diagrams in Fig.~\ref{fig:vertex}(c) with the dressed phonon propagator replaced by the BCS interaction or the unrenormalized phonon propagator. 
In the unrenormalized Migdal approximation, the expression for the self-energy reduces to
\begin{align}
\hat{\Sigma}^{\rm uMig}(t,t')=ig^2D_0(t,t')\hat{\sigma}_3\hat{G}_{\rm imp}(t,t')\hat{\sigma}_3,
\end{align} 
and we have 
\begin{align}
&\hat{\Gamma}(t,t';t'')=\hat{\Gamma}^{(0)}(t,t';t'')+ig^2D_0(t,t')\hat{\Lambda}(t,t';t'').\label{eq:vetex_uMig}
\end{align}
This equation is illustrated in Fig.~\ref{fig:vertex}(b).
Thus the 3rd and 4th diagrams in Fig.~\ref{fig:vertex}(c) 
represent the feedback from the phonon dynamics and have not been taken into account in the previous papers on collective modes.

\section{Results}

\subsection{Dynamical pair susceptibility and collective amplitude modes}
We now discuss the behavior of the pair susceptibility 
$\chi_{\rm pair}$ in the strongly-coupled SC. Figure ~\ref{fig:suscep_pair}(a)(b) displays $\chi^R_{\rm pair}(t)$.  
Also plotted is the bubble contribution $\chi^R_{0,{\rm pair}}(t)$, which is obtained by approximating $\hat{\Gamma}\approx\hat{\sigma}_1\delta_{\mathcal C} (t'',t)\delta_{\mathcal C} (t'',t')$.
While 
$\chi^R_{0,\text{pair}}(t)$ damps quickly (within $t\sim1/W$), $\chi^R_{\rm pair}(t)$ exhibits long-lived oscillations. Since $\chi_{0, {\rm pair}}$ only includes the contribution from the single-particle excitations, 
this 
indicates that the long-lived oscillations result from collective excitations.
A further finding is that, 
in contrast to the BCS dynamics, 
the oscillation contains multiple modes,
 which becomes more evident 
as we increase the el-ph coupling. 

We can capture the nature of the collective modes by comparing in Fig.~\ref{fig:suscep_pair}(c)(d) 
the electron spectrum 
$A(\omega)=-\frac{1}{\pi} G^R(\omega)$, the phonon spectrum $B(\omega)=-\frac{1}{\pi} D^R(\omega)$ 
and $-{\rm Im}\chi^R_{\rm pair} (\omega)=-{\rm Im}\int^{t_{\rm max}}_0 dt \chi^R_{\rm pair} (t) e^{i\omega t}$ with $t_{\rm max}=200$.   
First, we note that the strong el-ph coupling makes the gap edge smooth unlike in the BCS theory, and we define the gap size by the energy where $A(\Delta_{\rm SC})=N(0)$. We also note that, when the renormalized phonon frequency is comparable to the SC gap, the strong el-ph coupling leads to a highly asymmetric renormalized phonon spectrum in the SC state with a sharp peak below the SC gap, see Fig.~\ref{fig:suscep_pair}(c)(d) and Appendix~\ref{sec:phonon}. 
In the normal state, on the other hand, the phonon spectrum exhibits 
an almost symmetric single peak with a low-energy tail, and 
the renormalized phonon frequency is softened by the el-ph coupling.\cite{Hewson2004,Murakami2015} These features in the phonon spectra have indeed been 
experimentally observed in some strongly-coupled SCs,\cite{Axe1973,Axe1975,Kawano1996,Shapiro1997,Weber2008} and theoretically explained as an effect of the phonon self-energy \cite{Allen1997} (phonon anomaly). 

\begin{figure}[t]  
     \centering
     \vspace{-0.4cm}
  \includegraphics[width=85mm]{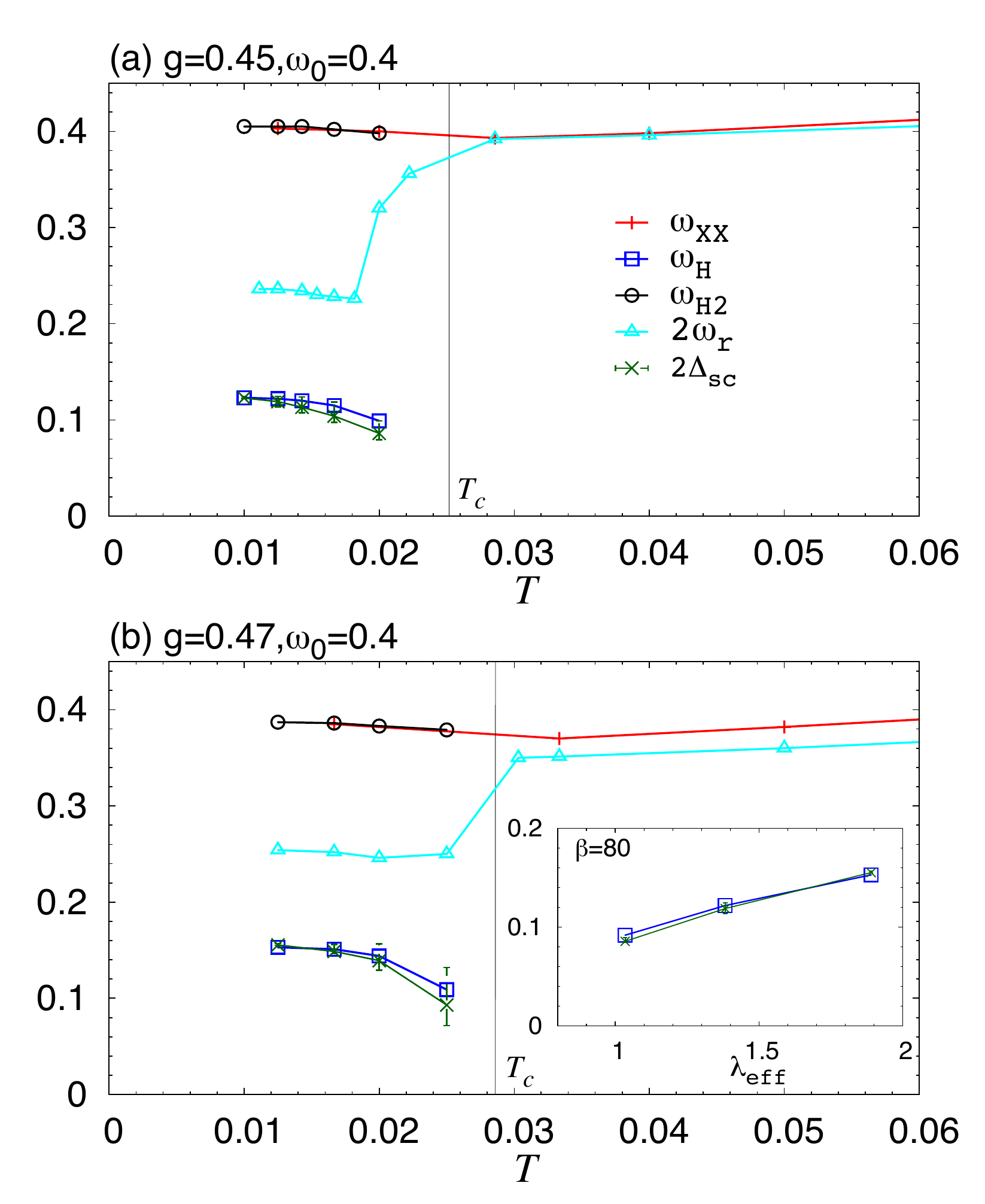}
  \caption{ Characteristic energies against temperature ($T$) at $g=0.45, \omega_0=0.4$ (a) and $g=0.47, \omega_0=0.4$ (b). 
Vertical lines indicate $T_c$. The inset shows the el-ph coupling ($\lambda_{\rm eff}$) dependence of $\omega_H$ and $2\Delta_{\rm SC}$ at $\beta=80$.}
  \label{fig:summary_energy_modes_w0.4}
  \end{figure}

Figure.~\ref{fig:suscep_pair}(c)(d) show that there exist two different modes in $-{\rm Im} \chi^R_{\rm pair}(\omega)$ at frequencies we call 
$\omega_{\rm H}$ and $\omega_{\rm H2}$ ($\omega_{\rm H}<\omega_{\rm H2}$). 
The lower-frequency peak is always located  
near the SC gap energy ($\omega/2\simeq\Delta_{\rm SC}$), and this also holds 
as we approach the 
BCS regime, see the inset of Fig.~\ref{fig:summary_energy_modes_w0.4}(b). We can thus identify this mode as the amplitude (Higgs) mode of the strongly-coupled SC. In other words, the BCS relation, $\omega_{\rm H}=2\Delta_{\rm SC}$, is found to hold to a good approximation 
even when the el-ph coupling is strong and the phonon energy is comparable to the gap.  
The higher-frequency mode at $\omega_{\rm H2}$ [Fig.~\ref{fig:suscep_pair}(c)(d)], on the other hand, is a new collective amplitude mode. This mode becomes more prominent as the el-ph coupling increases and is absent both in the BCS \cite{Anderson1958,Volkov1974,Kulik1981,Littlewood1981,Littlewood1982,Barankov2006,Yuzbashyan2006,Papenkort2007,Papenkort2008,Schnyder2011,Krull2014,Tsuji2015,Benfatto2014,Benfatto2015,Capone2015} and unrenormalized Migdal analyses.\cite{Kemper2015,Sentef2015}
These facts suggest that it does not exist in the weak-coupling regime, where the BCS treatment should be justified, and originates from the phonon dynamics.
However, $\omega_{\rm H2}$ does not have a simple relation 
with the renormalized phonon frequency $\omega_r$, which is defined as the peak position in the phonon spectrum.

In order to obtain a full picture, 
we plot the mode energies 
against $T$ in Fig.~\ref{fig:summary_energy_modes_w0.4}. 
As for the Higgs amplitude mode, we can see that the relation $\omega_{\rm H}\simeq 2\Delta_{\rm SC}$ 
is indeed robust for the whole region of $T$ studied here.\footnote{The uncertainty in the gap size is estimated from the peak position of $A(\omega)$ and the position where $A(\omega)=N(0)/2$ in Fig.~\ref{fig:summary_energy_modes_w0.4}.} 
We consider that the relation between $\omega_{\rm H}$ and $2\Delta_{\rm SC}$ is not obvious in the strongly-coupled regime. 
This is because, in principle, the Higgs mode can hybridize with other collective modes. One example is the Higgs mode in a coexistence region of SC and charge order, where the hybridization with the amplitude mode of the charge order can push the Higgs mode below the SC gap.\cite{Littlewood1981,Benfatto2014} In the present case, as demonstrated below, the Higgs mode hybridizes with the $\omega_{\rm H2}$ mode, 
which makes it slightly softened, but this effect is relatively small for $\lambda_{\rm eff}\lesssim2$, so that $\omega_{\rm H}$ remains close to $2\Delta_{\rm SC}$. 
One important consequence of the relation $\omega_{\rm H}\simeq2\Delta_{\rm SC}$ is that
the damping channel to quasi-particles remains small, which retains 
the amplitude Higgs mode long-lived.
This applies especially at low temperatures, where the gap edge is sharp enough and energy of quasi-particle excitations are lower-bounded at $2\Delta_{\rm SC}$. As we increase the temperature toward $T_c$, the quasi-particle lifetime from the strong el-ph coupling decreases and the gap edge becomes more smooth.
Hence the quasi-particle excitation is not strictly lower-bounded at $2\Delta_{\rm SC}$, which should lead to shorter lifetime of the Higgs oscillation.
Detailed analysis of the damping of the Higgs mode will be shown elsewhere.\cite{Murakami_in_prep}
In addition, we note that a possible relaxation channel from the Higgs mode into two phonons is restricted due to the suppression of the phonon spectral weight in the low-energy regime, 
which is associated with the phonon anomaly.

Now we turn to the $\omega_{\rm H2}$ mode.
Both the $\omega_\text{H}$ and $\omega_{\rm H2}$ modes are absent 
in the dynamical pair susceptibility in the normal state. On the other hand,
the latter mode is 
closely related to the coherent phonon oscillation that
persists at $T>T_c$.
In Fig.~\ref{fig:summary_energy_modes_w0.4}, we display $\omega_{\rm XX}$, the frequency of coherent oscillations
in the response of $\langle XX \rangle$ (i.e., the fluctuation of the phonon displacement) after a small hopping quench. 
We can see that $\omega_{\rm XX}$ coincides with 
$\omega_{\rm H2}$ in the SC phase, 
which indicates that the oscillations in these two different susceptibilities originate from the same collective mode.  
Hence this mode
intertwines both the phonon dynamics and superconducting amplitude oscillations. 
With decreasing temperature, 
$\omega_{\rm XX}$ softens in the normal phase, while it hardens in the SC. 
If the $\omega_{\rm H2}$, or $\omega_{\rm XX}$, mode were merely a coherent phonon mode,
the frequency would be equal to $2\omega_r$, where $\omega_r$ is the renormalized
phonon frequency defined as the position of the dominant peak in the phonon spectrum, see Appendix~\ref{sec:phonon}.
The factor $2$ appears because 
the present excitation does not induce any average phonon displacement.
\footnote{Our excitation protocols are homogeneous and do not change the total particle number. Hence no lattice distortion is induced, $\langle X \rangle=0$, and only symmetric dynamics is allowed for the statistical distribution for the lattice displacement. Hence 2$\omega_r$ is most relevant for the dynamics.}
This naive expectation ($\omega_{\rm XX}=2\omega_r$) is satisfied in the normal state with not too strong el-ph couplings, while in the SC phase 
$\omega_{\rm H2}$ $(=\omega_{\rm XX})$ drastically deviates from $2\omega_r$, see Fig.~\ref{fig:summary_energy_modes_w0.4}.

\subsection{Diagrammatic analysis}

\begin{figure}[t]  
     \centering
     \vspace{-0.2cm}
  \includegraphics[width=90mm]{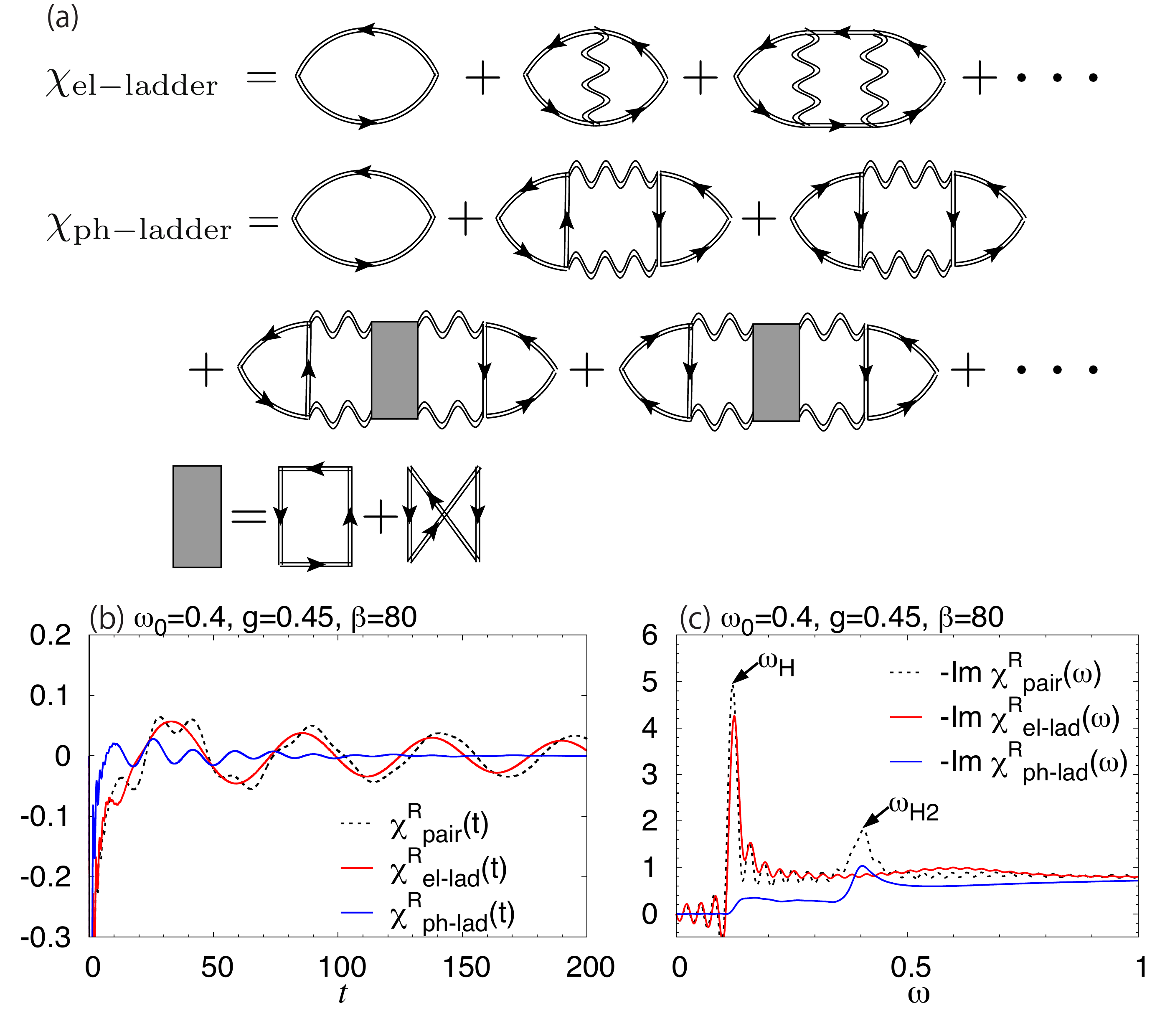}
  \caption{(a) Diagrammatic expression for $\chi_{\rm el-ladder}$ and $\chi_{\rm ph-ladder}$ with the electron-mediated phonon-phonon interaction (shaded box).
   Comparison of contributions from different sets of diagrams for
  $\chi(t)$ (b) and $-{\rm Im}\chi(\omega)$ (c) with $g=0.45,\omega_0=0.4, \beta=80$. 
}
  \label{fig:Chi_diagram_origin}
  \end{figure}

In this section, we address 
 (1) the effect of the phonon dynamics on the amplitude Higgs mode, and 
(2) the origin of the discrepancy between $\omega_{\rm H2}$ and $2\omega_r$ in the SC. 
To gain some insights, we  
evaluate the contributions from 
certain subsets of the diagrams for $\chi_{\rm pair}$. The first subset is $\chi_{\rm el-ladder}$, which is illustrated in Fig.~\ref{fig:Chi_diagram_origin}(a).
In the BCS and unrenormalized Migdal approximations, where the non-equilibrium dynamics of the phonons is neglected, the vertex $\hat{\Gamma}$ includes the first two diagrams in Fig.~\ref{fig:vertex}(c), which leads to ladder diagrams with electron legs. 
 Hence we can regard $\chi_{\rm el-ladder}$ as the contribution without phonon dynamics.
 Indeed, this subset is evaluated by considering the time evolution with
 \begin{subequations}
  \begin{align}
\hat{\Sigma}(t,t')&=ig^2D_{\rm imp}^{\rm eq}(t,t')\hat{\sigma}_3\hat{G}_{{\rm imp}}(t,t')\hat{\sigma}_3,\\
\Pi(t,t')&=-ig^2{\rm tr} [\hat{\sigma}_3\hat{G}_{\rm imp}^{\rm eq}(t,t')\hat{\sigma}_3\hat{G}_{\rm imp}^{\rm eq}(t',t)].
\end{align}
\end{subequations}
Here, ``eq" indicates that the functions are fixed to their equilibrium value. 
  On the other hand, the 3rd and 4th diagrams for the vertex in Fig.~\ref{fig:vertex}(c) only appear in the self-consistent Migdal approximation, 
 and thus 
 represent the effect of the phonon dynamics. By eliminating the 2nd diagram in the vertex in Fig.~\ref{fig:vertex}(c), we obtain a set of diagrams for the pair susceptibility, which is illustrated as $\chi_{\rm ph-ladder}$ in Fig.~\ref{fig:Chi_diagram_origin}(a) and represents the contribution from the phonon dynamics. 
We can evaluate $\chi_{\rm ph-ladder}$ by computing the time evolution with
 \begin{subequations}\label{eq:sig_mig}
  \begin{align}
\hat{\Sigma}(t,t')&=ig^2D_{\rm imp}(t,t')\hat{\sigma}_3\hat{G}_{\rm imp}^{\rm eq}(t,t')\hat{\sigma}_3,\\
\Pi(t,t')&=-ig^2{\rm tr} [\hat{\sigma}_3\hat{G}_{\rm imp}(t,t')\hat{\sigma}_3\hat{G}_{\rm imp}(t',t)].
\end{align}
\end{subequations}
 
 In Fig.~\ref{fig:Chi_diagram_origin}(b)(c), we plot 
 $\chi^R_{\rm pair}$, $\chi^R_{\rm el-ladder}$ and $\chi^R_{\rm ph-ladder}$
 in the time and frequency domain. 
It turns out that each of $\chi^R_{\rm el-ladder}(t)$ and $\chi^R_{\rm ph-ladder}(t)$ 
exhibits oscillations with a single characteristic frequency, 
which agrees well with $\omega_{\rm H}$ and $\omega_{\rm H2}$, respectively. 
Hence $\omega_{\rm H}$ and $\omega_{\rm H2}$ are mainly determined by the processes 
represented by $\chi_{\rm el-ladder}$ and $\chi_{\rm ph-ladder}$, respectively.

As for question (1), even though the response without phonon dynamics mainly sets the energy scale of the Higgs mode, we do observe effects from the phonon dynamics, in the form of a phase shift in the Higgs oscillation between $\chi^R_{\rm pair}(t)$ and $\chi^R_{\rm el-ladder}(t)$, and also in the form of an overestimation of $\omega_{\rm H}$ in the latter approximation by several percent, see Fig.~\ref{fig:Chi_diagram_origin}(b)(c). In addition, larger intensity in $\chi^R_{\rm pair}(\omega)$ at $\omega_{\rm H}$ than in $\chi^R_{\rm el-ladder}(\omega)$ is consistent with that the damping of the Higgs oscillation should be slower in the former because the softened  $\omega_{\rm H}$ value leads to suppression of available quasi-particle relaxation channels.
These differences can be attributed to the remaining diagrams in $\chi_{\rm pair}$, which are not included in $\chi_{\rm el-ladder}$ and $\chi_{\rm ph-ladder}$. 
These diagrams hybridize electron ladders and phonon ladders, and the decrease of $\omega_{\rm H}$ estimated from the $\chi_{\rm el-ladder}$ can be ascribed to an effect of the hybridization between the Higgs mode in $\chi_{\rm el-ladder}$ and the phonon-origin mode in $\chi_{\rm ph-ladder}$.

As for question (2), we first note that $2\omega_r$-oscillations are expected from the two parallel phonon propagators in the 2nd and 3rd diagrams for $\chi_{\rm ph-ladder}$. 
Therefore, the notable hardening of $\omega_{\rm H2}$ from $2\omega_r$ in the SC is attributed to what we can call the ``electron-mediated phonon-phonon interactions" [the shaded rectangle in Fig.~\ref{fig:Chi_diagram_origin}(a)], while in the normal phase this effect is weaker. 

\begin{figure}[t]
     \centering
     \vspace{-0.2cm}
  \includegraphics[width=100mm]{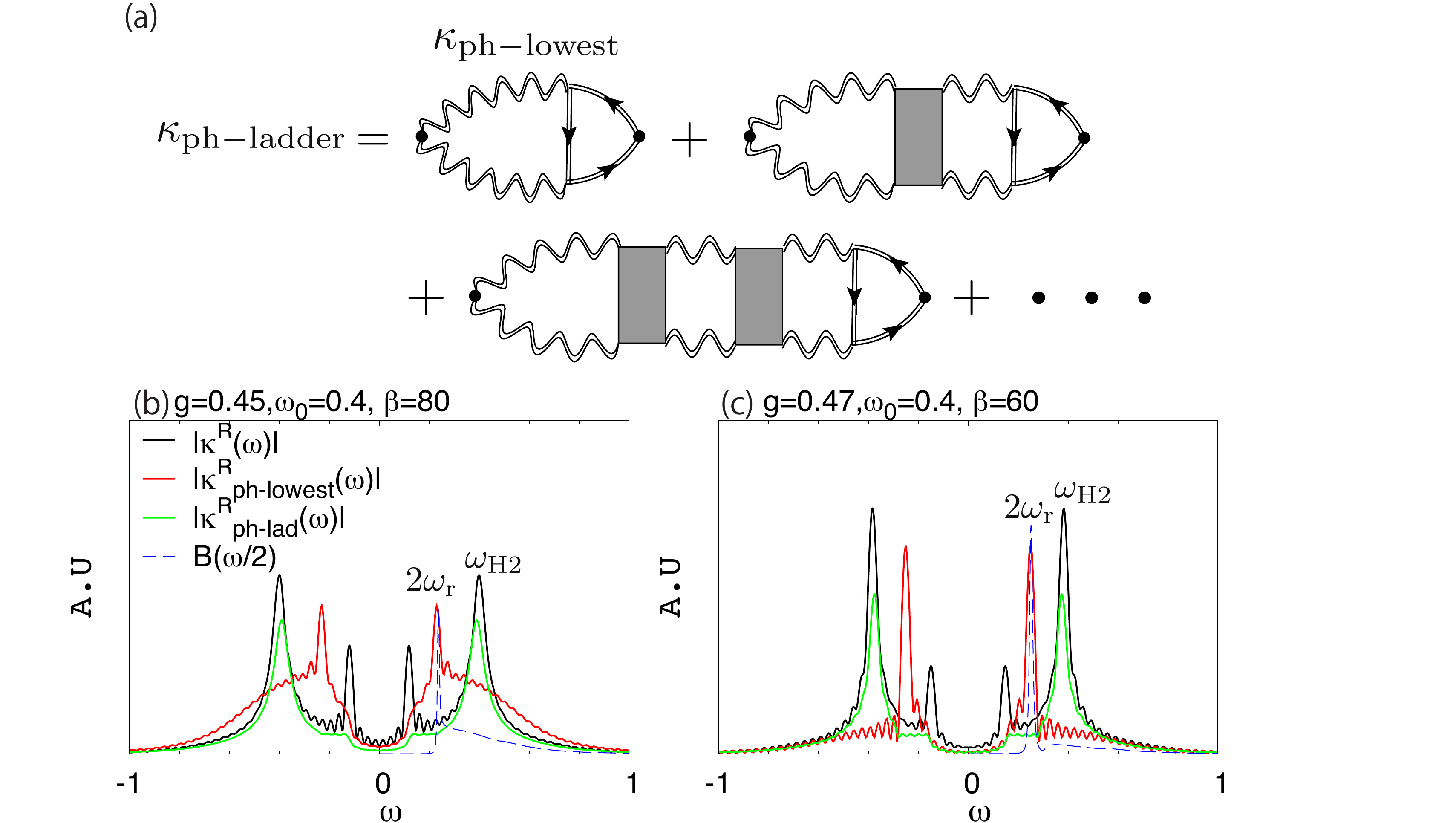}
  \caption{(a) Diagrammatic expression for $\kappa_{\rm ph-ladder}$ and $\kappa_{\rm ph-lowest}$. The data at $t\in[0,200]$ is used. (b)(c) Comparison of $\kappa^R(\omega)$ evaluated from different sets of diagrams and the phonon spectrum, $B(\omega)$, for $g=0.45,\;\omega_0=0.4,\;\beta=80$ (b) and $g=0.47,\;\omega_0=0.4,\;\beta=60$ (c).}
 \label{fig:lowest_ph_cont}
\end{figure}

We can also confirm the effect of the phonon-phonon (ph-ph) interaction mentioned above in another susceptibility.
Here, we focus on  $\kappa^R(t)\equiv-i\theta(t)\langle[XX(t),B_{\bf 0}]\rangle$ (response of $XX$ against the external pair field). 
This quantity can be expressed in terms of $\Omega(t,t';t'')$ defined in Eq.~(\ref{eq:Omega}) in Appendix~\ref{sec:chi}.
 Now, we evaluate a subset of diagrams, $\kappa_{\rm ph-ladder}$, which 
corresponds to $\chi_{\rm ph-ladder}$ in that the dynamics is described by Eq.~(\ref{eq:sig_mig}). 
The diagrammatic expression for $\kappa_{\rm ph-ladder}$ is shown in Fig.~\ref{fig:lowest_ph_cont}(a). Here one can note that the 1st diagram in $\kappa_{\rm ph-ladder}$ corresponds to the 2nd and 3rd ones in $\chi_{\rm ph-ladder}$. In the following, we denote the contribution from the 1st diagram as $\kappa_{\rm ph-lowest}$.
 In Fig.~\ref{fig:lowest_ph_cont}(b)(c) we compare all 
four of the $\kappa^R(\omega)$ from the full dynamics in the self-consistent Migdal approximation, $\kappa^R_{\rm ph-lowest}(\omega)$, $\kappa^R_{\rm ph-ladder}(\omega)$, and the phonon spectrum $B(\omega/2)$. 
 From the result one finds that $\kappa^R_{\rm ph-lowest}(\omega)$ indeed 
exhibits a peak at $2\omega_r$, which deviates from $\omega_{\rm H2}$. However, if one takes account of the effect of the ph-ph interaction as in $\kappa^R_{\rm ph-ladder}(\omega)$, there emerges a peak around $\omega_{\rm H2}$.

 \begin{figure}[t]
     \centering
     \vspace{-0.6cm}
  \includegraphics[width=90mm]{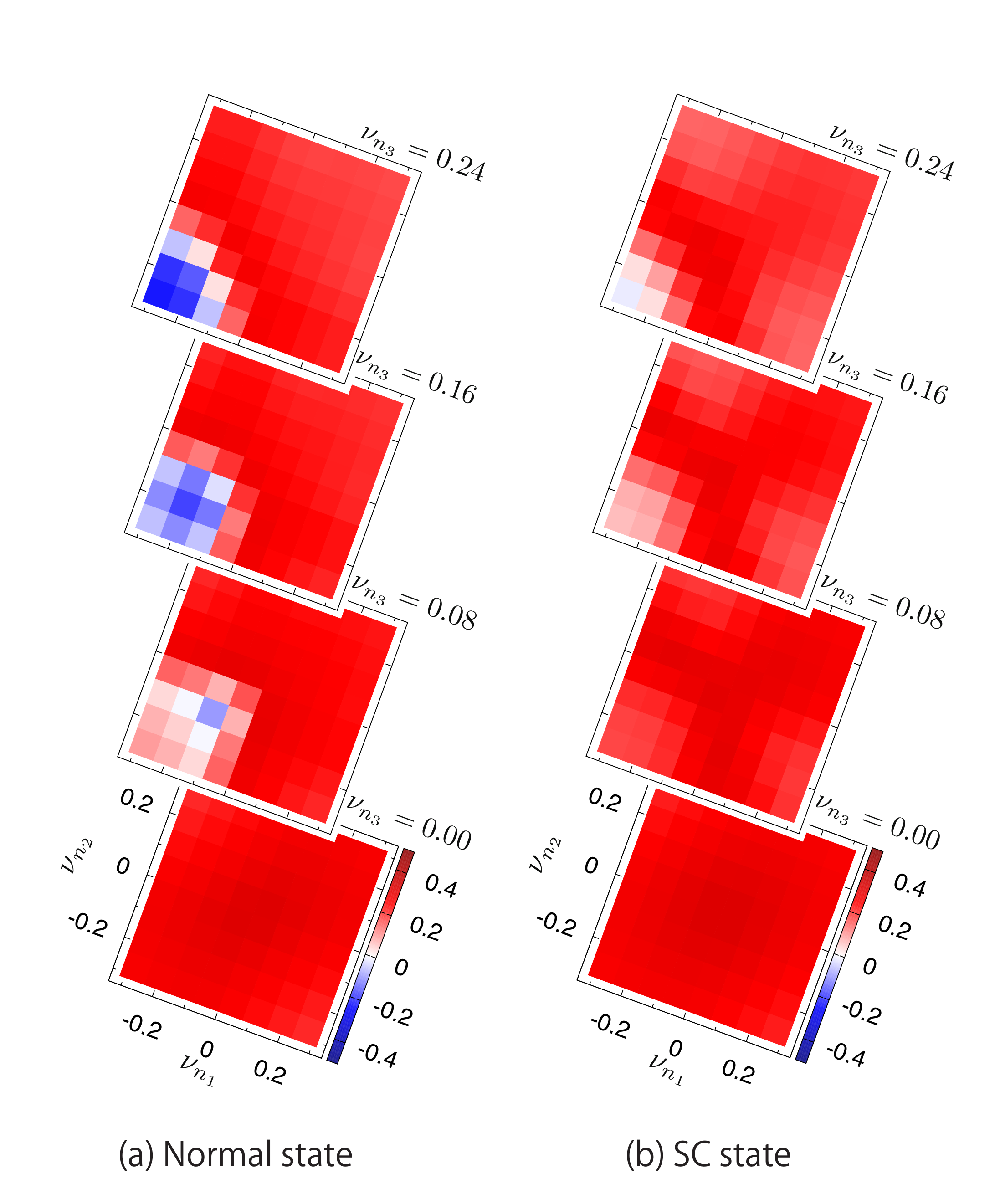}
  \caption{$I^{\rm M}_{\rm ph}(i\nu_{n_1},i\nu_{n_2},i\nu_{n_3})/g^4$ for various values of $\nu_{n_3}$ in the normal state (a) and in the SC state (b) for $g=0.45,\omega_0=0.4,\beta=80$. For the normal state, we suppress SC by hand.}
 \label{fig:ph_int_bcs_b400}
\end{figure}
The different effects of the ph-ph interaction in the SC and normal states can be attributed to the difference in its behavior in these phases.
The expression for the ph-ph interaction on the KB contour is 
\begin{align}
&I_{\rm ph}(t_1,t_2,t_3,t_4)\equiv\nonumber\\
& \frac{g^4}{N}\sum_{\bf k} \bigl\{{\rm tr}[\hat{\sigma}_3\hat{G}(t_1,t_2)\hat{\sigma}_3\hat{G}_{\bf k}(t_2,t_4)\hat{\sigma}_3\hat{G}(t_4,t_3)\hat{\sigma}_3\hat{G}_{\bf k}(t_3,t_1)]\nonumber\\
&\;\;\;\;\;+{\rm tr}[\hat{\sigma}_3\hat{G}(t_1,t_2)\hat{\sigma}_3\hat{G}_{\bf k}(t_2,t_3)\hat{\sigma}_3\hat{G}(t_3,t_4)\hat{\sigma}_3\hat{G}_{\bf k}(t_4,t_1)]\bigl\}.
\end{align}
In particular, the Matsubara components are
\begin{align}
&I^{\rm M}_{\rm ph}(\tau_1,\tau_2,\tau_3)\equiv\nonumber\\
& \frac{g^4}{N}\sum_{\bf k} \bigl\{{\rm tr}[\hat{\sigma}_3\hat{G}(\tau_1-\tau_2)\hat{\sigma}_3\hat{G}_{\bf k}(\tau_2)\hat{\sigma}_3\hat{G}(-\tau_3)\hat{\sigma}_3\hat{G}_{\bf k}(\tau_3-\tau_1)]\nonumber\\
&\;\;\;\;\;+{\rm tr}[\hat{\sigma}_3\hat{G}(\tau_1-\tau_2)\hat{\sigma}_3\hat{G}_{\bf k}(\tau_2-\tau_3)\hat{\sigma}_3\hat{G}(\tau_3)\hat{\sigma}_3\hat{G}_{\bf k}(-\tau_1)]\bigl\},
\end{align}
and 
\begin{align}
&I^{\rm M}_{\rm ph}(i\nu_{n_1},i\nu_{n_2},i\nu_{n_3})\equiv\nonumber\\
&\;\;\;\;\int^\beta_0 d\tau_1d\tau_2d\tau_3 e^{i\nu_{n_1}\tau_1}e^{i\nu_{n_2}\tau_2}e^{i\nu_{n_3}\tau_3}I^{\rm M}_{\rm ph}(\tau_1,\tau_2,\tau_3),
\end{align}
where $\nu_{n_\alpha}=2n_{\alpha}{\pi}/{\beta}$.
Now, in order to clarify the difference in the ph-ph interaction in the normal and SC phases, we directly evaluate $I^{\rm M}_{\rm ph}(i\nu_{n_1},i\nu_{n_2},i\nu_{n_3})$.
In Fig.~\ref{fig:ph_int_bcs_b400}, we show the results for the normal and SC cases for $g=0.45,\;\omega_0=0.4,\;\beta=80$.  
In order to obtain the result for the normal state, we suppress SC by hand.
First we note that $I^{\rm M}_{\rm ph}(i\nu_{n_1},i\nu_{n_2},i\nu_{n_3})$ is real. 
In the normal state, the ph-ph interaction strongly depends on the Matsubara frequency and has a clear sign change. 
On the other hand, the SC phase has a drastically different 
behavior: In the frequency regime comparable to the SC gap, the frequency dependence becomes much weaker and the sign change almost disappears.
This allows us to approximate $I^{\rm M}_{\rm ph}(i\nu_{n_1},i\nu_{n_2},i\nu_{n_3})$ by a constant in the SC state. 
From a  comparison with diagrams that appear in the perturbation expansion for a simple phonon model with an anharmonic term, $H^{\rm eff}_{\rm ph}=\omega_{\rm ph} a^\dagger a+I_4 X^4$, it turns out that an approximate constant $I^{\rm M}_{\rm ph}(i\nu_{n_1},i\nu_{n_2},i\nu_{n_3})$ corresponds to the case of $I_4>0$. Since the anharmonic term makes the potential steeper, the frequency of the coherent oscillations 
increases for $I_4>0$. 
This analysis is indeed consistent with our observation of the hardening from $2\omega_r$ to $\omega_{\rm H2}$ in the SC phase. In the normal state, it is expected that the cancellation from the sign change 
in the frequency dependence 
reduces this effect.
 \begin{figure*}[t]
     \centering
  \includegraphics[width=180mm]{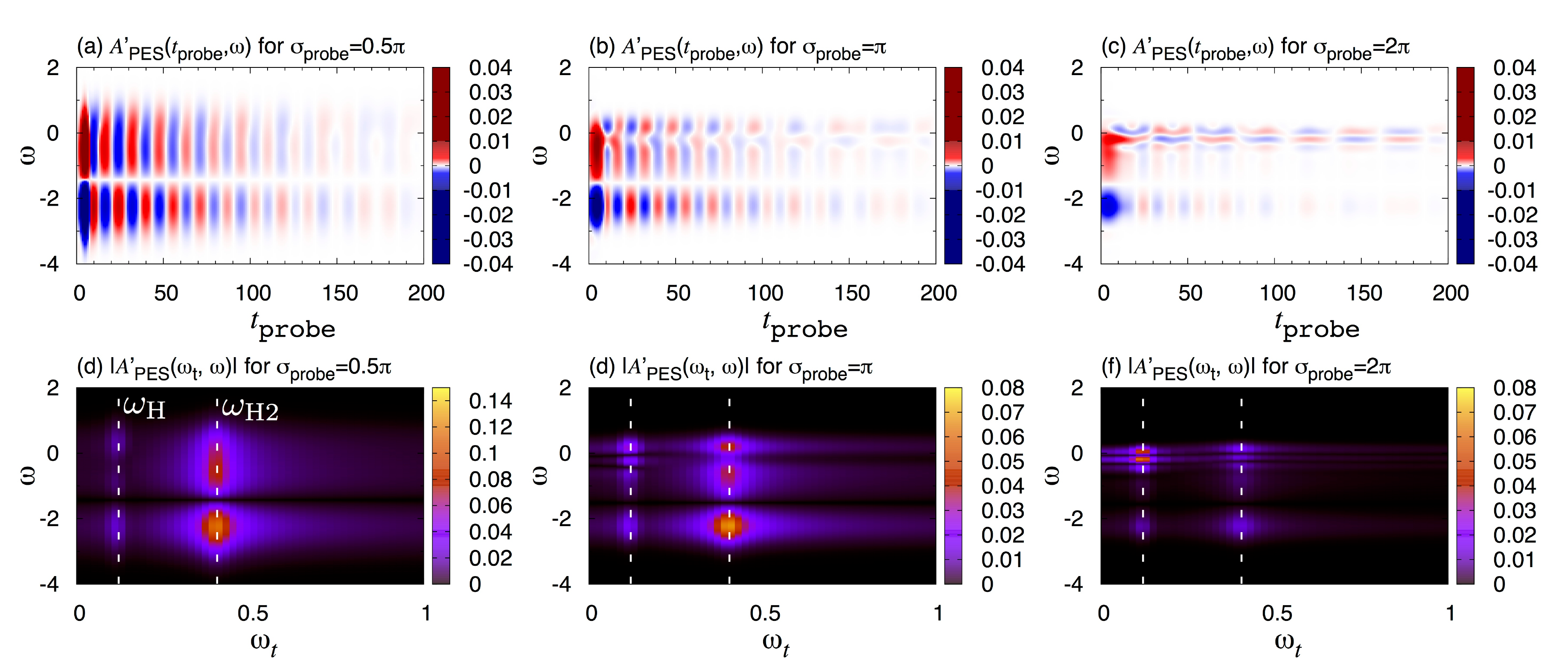}
  \caption{(a)(b)(c) $A'_{\rm PES} (t,\omega)$ against $t$ and $\omega$ for various values of $\sigma_{\rm probe}$ for $g=0.45,\omega_0=0.4,\beta=80$. (d)(e)(f) $|A'_{\rm PES}(\omega_t,\omega)|$ against $t$ and $\omega$ for various values of $\sigma_{\rm probe}$ for $g=0.45,\omega_0=0.4,\beta=80$. The white vertical lines in (d)(e)(f) indicate $\omega_{\rm H}$ and $\omega_{\rm H2}$. The condition for the pump and probe is $t_{\rm pump}=5.0$, $\sigma_{\rm pump}=1.0$ and $t_c=25$.}
  \label{fig:tARPES_spect_tpump_g0.45w0.4b80}
\end{figure*}

\subsection{Time-resolved photoemission spectroscopy}

Although the dynamical pair correlation is not a direct observable in
experiments, here we discuss that these modes can be observed in pump-probe
spectroscopy measurements.
We focus on the dynamics of the spectral function observed in time-resolved photoemission spectroscopy (tr-PES), \cite{Freericks2009}
\begin{align}
&A_{\rm PES}(t_{\rm probe},\omega)\equiv\nonumber\\
 &\frac{1}{\pi} {\rm Im} \int dt dt' s(t-t_{\rm probe})s(t'-t_{\rm probe})e^{i\omega(t-t')}G^<(t,t'). 
 \end{align}
Here $t_{\rm probe}$ is the center of the probe pulse and $s(t)$ is its envelope, for which we use a Gaussian with sufficiently large cutoff time $t_c$, i.e. $s(t)=\frac{1}{\sqrt{2\pi}\sigma_{\rm probe}}\exp(-\frac{t^2}{2\sigma_{\rm probe}^2})\theta(t_c-|t|)$.
We note that a previous study with equilibrium phonons has pointed out the possibility of detecting the amplitude mode in the time-resolved photoemission signal near the Fermi level. \cite{Kemper2015} 
Here we focus on a wider energy range and the new amplitude mode.
The pump is mimicked by a modulation of the hopping, 
\begin{align}
v(t)=v_{0}+\delta v\exp\left[-\frac{(t-t_{\rm pump})^2}{2\sigma_{\rm pump}^2}\right],
\end{align}
where $\sigma_{\rm pump}$ and $t_{\rm pump}$ respectively denote the width and the center of the pump pulse.  
This type of pump can be effectively realized with a strong laser through an effective band renormalization\cite{Tsuji2011} or through light-induced lattice distortions.\cite{Sentef2015,Subedi2011,Forst2011a}
In the following, we choose $t_{\rm pump}=5.0$, $\sigma_{\rm pump}=1.0$, and $t_c=25$.
In Fig.~\ref{fig:tARPES_spect_tpump_g0.45w0.4b80}(a)(b)(c), we plot the difference between the spectra with and without a pump normalized by the pump strength $\delta v$,
\begin{align}
&A'_{\rm PES}(t_{\rm probe},\omega)\equiv\nonumber\\
&\lim_{\delta v\rightarrow 0}\frac{A_{\rm PES}(t_{\rm probe},\omega;\delta v)-A_{\rm PES}(t_{\rm probe},\omega;0)}{\delta v},
\end{align}
where we add the third argument for $A_{\rm PES}$, which indicates the strength of the pump. Clear oscillations are seen in a wide energy range for $\omega\alt 0$. 
One can see that for smaller $\sigma_{\rm probe}$ the resolution for $\omega$ decreases, while for $t_{\rm probe}$ it increases.
The Fourier transform along $t_{\rm probe}$, 
\begin{align}
A'_{\rm PES}(\omega_t,\omega)\equiv\int^{t_{\rm max}-t_c}_{t_c} dt A'_{\rm PES}(t,\omega)e^{i\omega_t t},
\end{align}
reveals that the dominant oscillations
are the $\omega_{\rm H}$ and $\omega_{\rm H2}$ components, see Fig.~\ref{fig:tARPES_spect_tpump_g0.45w0.4b80}(d)(e)(f).
These signals are visible in a wide energy range for $\omega\alt 0$ as bundles at the corresponding energies, especially near the band edge ($\omega\simeq W/2$) and the gap edge ($\omega\simeq\Delta_{\rm SC})$.

\begin{figure}[h]
     \centering
     \vspace{-0.cm}
  \includegraphics[width=85mm]{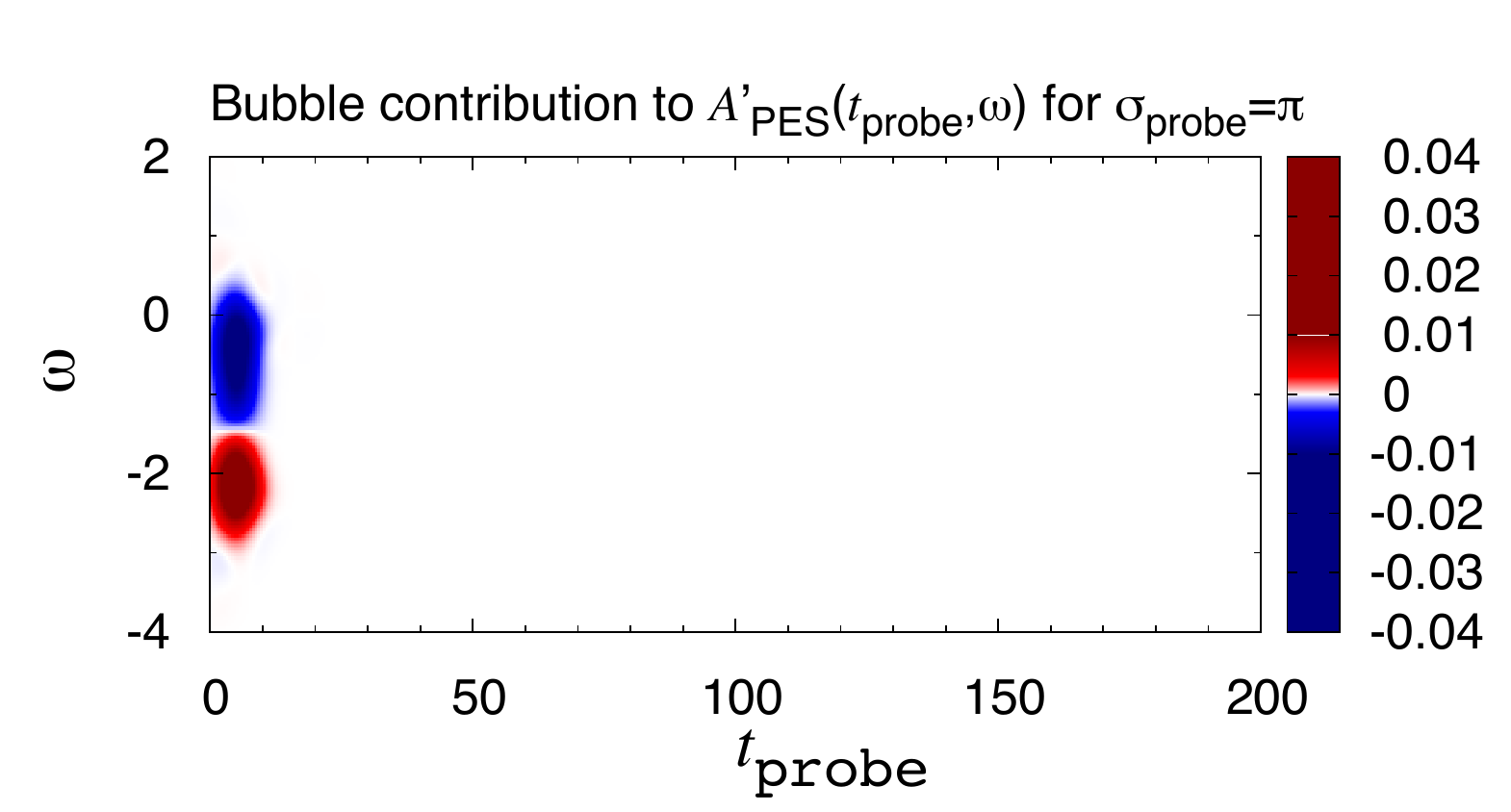}
  \caption{Contribution of the bubble diagrams to $A'_{\rm PES}(t_{\rm probe},\omega)$ at $g=0.45, \omega_0=0.4, \beta=80$. The condition for the pump and probe is $t_{\rm pump}=5.0$, $\sigma_{\rm pump}=1.0$, $t_c=25$ and $\sigma_{\rm probe}=\pi$. }
 \label{fig:test2}
\end{figure}

Finally, we demonstrate that the oscillations in the tr-PES spectra cannot be explained by single particle excitations, i.e. the contribution from the bubble diagram.
In the linear-response regime, we have to consider a quantity for the tr-PES spectrum,
\begin{align}
&\left.\frac{\delta_{\mathcal C} [\hat{G}_{\bf k}(t,t')]}{{\delta_{\mathcal C} [v(t'')}]}\right|_{v(t)=v}\nonumber\\
&=\int_{\mathcal C} d{t_1}d{t_2}\hat{G}_{\bf k}(t,t_1)\hat{\Gamma}_{{\rm hop},{\bf k}}(t_1,t_2;t'')\hat{G}_{\bf k}(t_2,t').
\end{align}
Following the same procedure as for the pair susceptibility, we obtain the expression for the vertex part as 
\begin{align}
\hat{\Gamma}_{{\rm hop},{\bf k}}(t,t';t'')=\hat{\Gamma}^{(0)}_{{\rm hop},{\bf k}}(t,t';t'')+\left.\frac{\delta_{\mathcal C} [\hat{\Sigma}(t,t')]}{{\delta_{\mathcal C} [v(t'')}]}\right|_{v(t)=v}.
\end{align}
Here, $\hat{\Gamma}^{(0)}_{{\rm hop},{\bf k}}(t,t';t'')=\frac{\epsilon_{\bf k}}{v}\hat{\sigma}_3\delta_{\mathcal C} (t'',t)\delta_{\mathcal C} (t'',t')$ and the second term is the vertex correction. The result for the tr-PES spectrum, $A'_{\rm PES}(t_{\rm probe},\omega)$, evaluated with the bubble contribution (without the vertex correction)
is displayed in Fig.~\ref{fig:test2}.  While during the pump it shows a similar behavior as the full dynamics in Fig.~\ref{fig:tARPES_spect_tpump_g0.45w0.4b80}, 
there is no oscillation after the pump. 
 Hence we again conclude that the oscillations in the photoemission spectra do originate from collective excitations, and we predict that they can be observed in pump-probe experiments in a wide range of $\omega$. 

\section{Conclusion}
We have investigated the properties of collective amplitude modes in strongly-coupled SC in the Holstein model using the non-equilibrium DMFT 
implemented with the self-consistent Migdal approximation. 
The BCS relation between the SC gap and the Higgs energy turns out to be robust beyond the BCS regime. 
Besides the Higgs mode, we have unraveled another amplitude mode involving the dynamics of the phonons. 
The frequency of this mode, higher than twice the renormalized phonon frequency in the superconducting phase, was shown to reflect a strong electron-mediated phonon-phonon interaction. 
We have also predicted that both collective modes should be observable as oscillations of the PES spectrum in a wide energy range after a strong laser pump. 
Even though the new mode involving the dynamics of the SC order parameter and the phonon dynamics has not yet been observed in real materials, this information would be helpful for searching for such a mode.
We stress that the Holstein model is a fundamental model describing the essential physics of electron-phonon systems with a local coupling. In addition, our approximate method, the non-equilibrium DMFT+ self-consistent Migdal approximation, is the non-equilibrium extension of the Migdal-Eliashberg theory, which has been successful in describing strongly-coupled conventional SCs, in the limit of infinite spatial dimensions. Hence we believe that the present study for a fundamental model with a fundamental approximation will be a milestone for further analyses for collective excitations beyond the BCS limit.
An interesting future direction is to study these collective modes in more realistic setups, such as multi-band systems\cite{Manske2013,Murotani2015,Schnyder2015}, 
models with local and nonlocal Coulomb interactions, and more general el-ph couplings. In order to deal with these setups, further development of impurity solvers and/or extension of the DMFT framework are required.
We also note that it has been recently pointed out that, depending on details of the pump excitation, there can be significant contributions from quasi-particle excitations to the third-harmonic generation with a strong THz excitation.\cite{Benfatto2015b} Therefore, systematic studies of how the contributions from quasi-particle excitations and collective excitations depend on the excitation protocols and observables would also be important future works.
  
 \acknowledgments 
  The authors wish to thank D. Golez, H. Strand and R. Matsunaga for fruitful discussions.
YM, NT, and HA wish to thank the Fribourg University for hospitality when the manuscript was written. 
HA is supported by a Grant-in-Aid for Scientific Research (Grant No. 26247057)  from MEXT and ImPACT project (No. 2015-PM12-05-01) from JST, while YM is supported by JSPS Research Fellowships for Young Scientists and Advanced Leading Graduate Course for Photon Science (ALPS). PW acknowledges support from FP7 ERC starting grant No. 278023. 
NT is supported by Grants-in-Aid for Scientific Research from JSPS (No. 25800192).

\bibliographystyle{prsty}
\bibliography{thesis}

\appendix
\section{Renormalized phonons} \label{sec:phonon}
\begin{figure}[t]
     \centering
   \includegraphics[width=60mm]{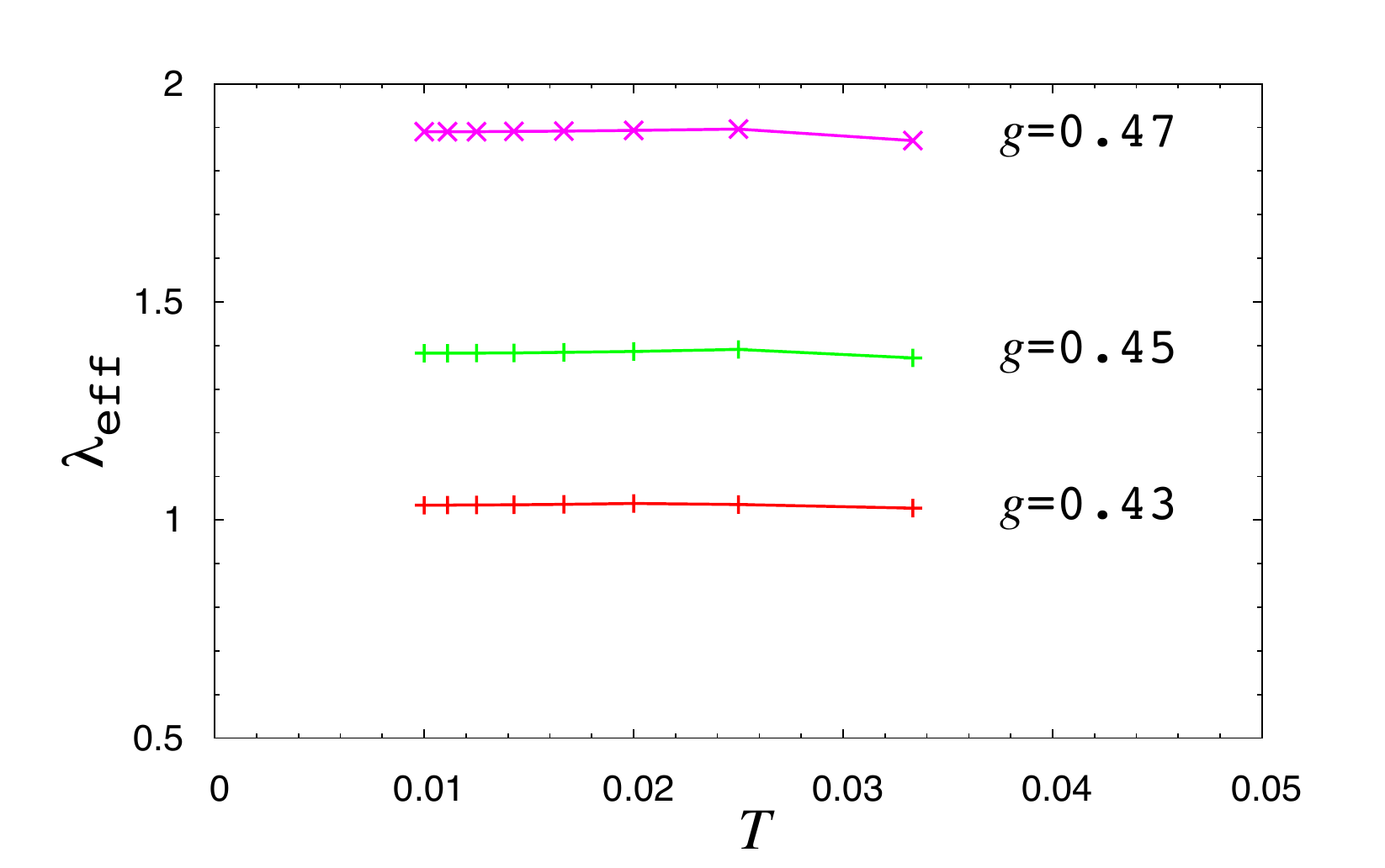}
  \caption{The dimensionless el-ph coupling $\lambda_{\rm eff}$ against 
$T$ for various values of $g$.}
  \label{fig:lam_eff_w0.4}
\end{figure}

Here we explain the dimensionless el-ph coupling for renormalized phonons and the behavior of the 
phonon spectrum in SC.
The dimensionless el-ph coupling is defined as
\begin{align}
\lambda_{\rm eff}&\equiv2\int^{\infty}_0 d\omega \frac{\alpha^2F(\omega)}{\omega},\\
\alpha^2F(\omega)&=N(0)g^2B(\omega),
\end{align}
where $N(0)$ is the DOS at the Fermi level, $B(\omega)=-\frac{1}{\pi} {\rm Im} D^R(\omega)$ and we obtain $\lambda=N(0)g^2D^M(i\nu_n=0)$.
Here the superscript $M$ indicates the Matsubara component.
So called strong-coupling superconductors correspond to cases of $\lambda_{\rm eff}\sim 1$.
The $\lambda_{\rm eff}$ for the parameters employed in the paper is show in Fig.~\ref{fig:lam_eff_w0.4}. 
In all cases, the temperature dependence is weak.

\begin{figure}[t]
     \centering
     \vspace{-0.cm}
   \includegraphics[width=90mm]{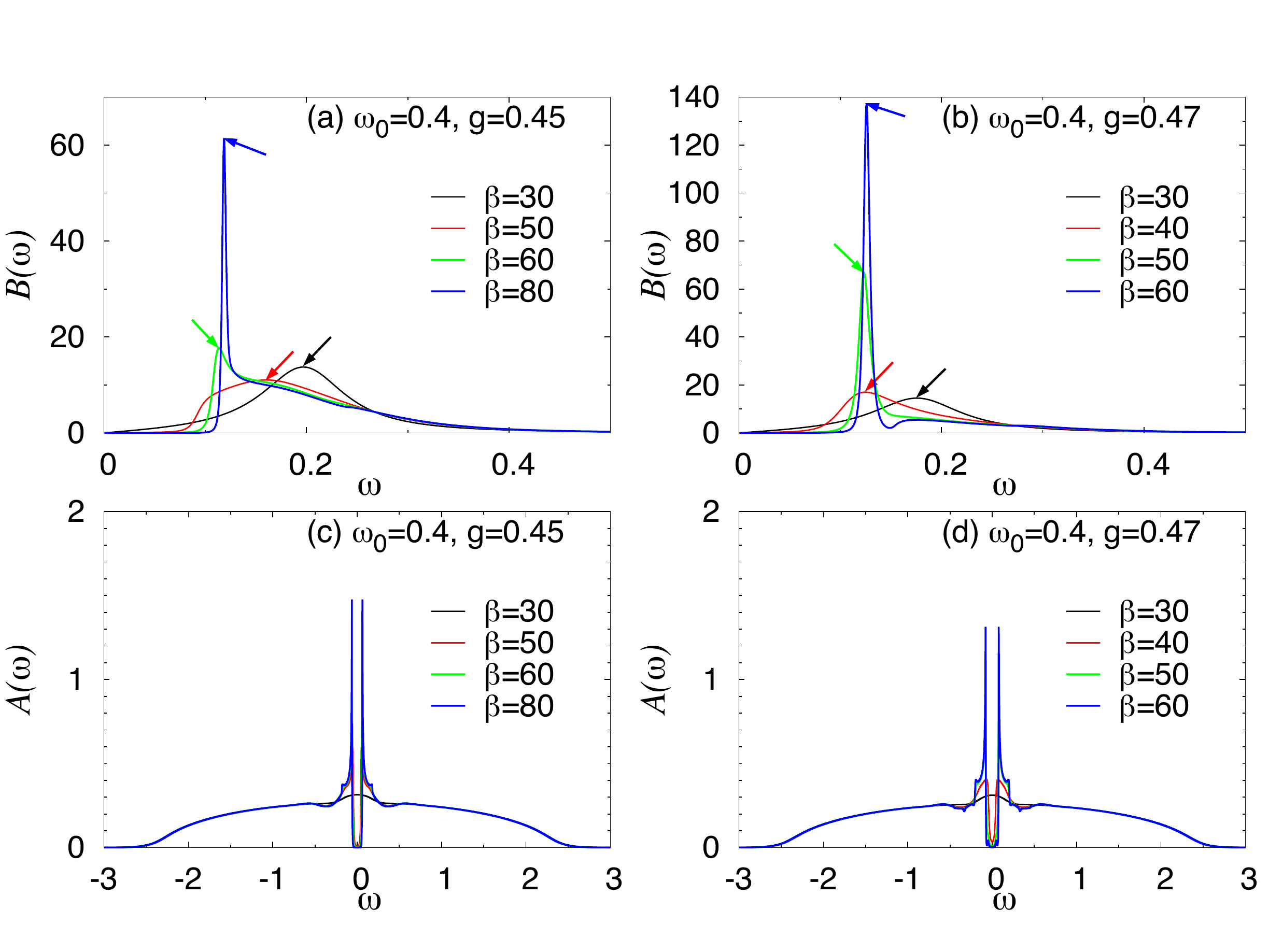}
  \caption{(a)(b) Phonon spectrum at various temperatures for $\omega_0=0.4,\;g=0.45$ (a) and  $\omega_0=0.4,\;g=0.47$ (b). Arrows indicate the peak positions ($\omega_r$). 
  (c)(d) Election spectrum at various temperatures for $\omega_0=0.4,\;g=0.45$ (c) and  $\omega_0=0.4,\;g=0.47$ (d).
  In all cases, the system is in the normal phase at $\beta=30$, while in the SC phase at other temperatures. }
  \label{fig:phonon_spectrum_Mig_w0.4}
\end{figure}

In Fig.~\ref{fig:phonon_spectrum_Mig_w0.4}, we show the detailed temperature dependence of the phonon spectrums for the cases corresponding to Fig.~\ref{fig:summary_energy_modes_w0.4} as well as the electron spectra for the whole energy range.
For both couplings, $\beta=30$ is in the normal state, and the phonon spectrums exhibit an almost symmetric structure around a peak at $\omega_r$.
In the SC phase, the SC gap develops as we decrease the temperature. 
At the same time, there occurs a drastic change of the phonon spectrum.
In particular, the spectral weight in the low energy regime is strongly suppressed and a sharp peak develops below the SC gap.
The former is attributed to the fact that scattering 
of phonons with quasi-particles is suppressed below the SC gap energy.

\section{Results for $\omega_0=0.2$}\label{sec:0.2}
\begin{figure}[t]
     \centering
   \includegraphics[width=90mm]{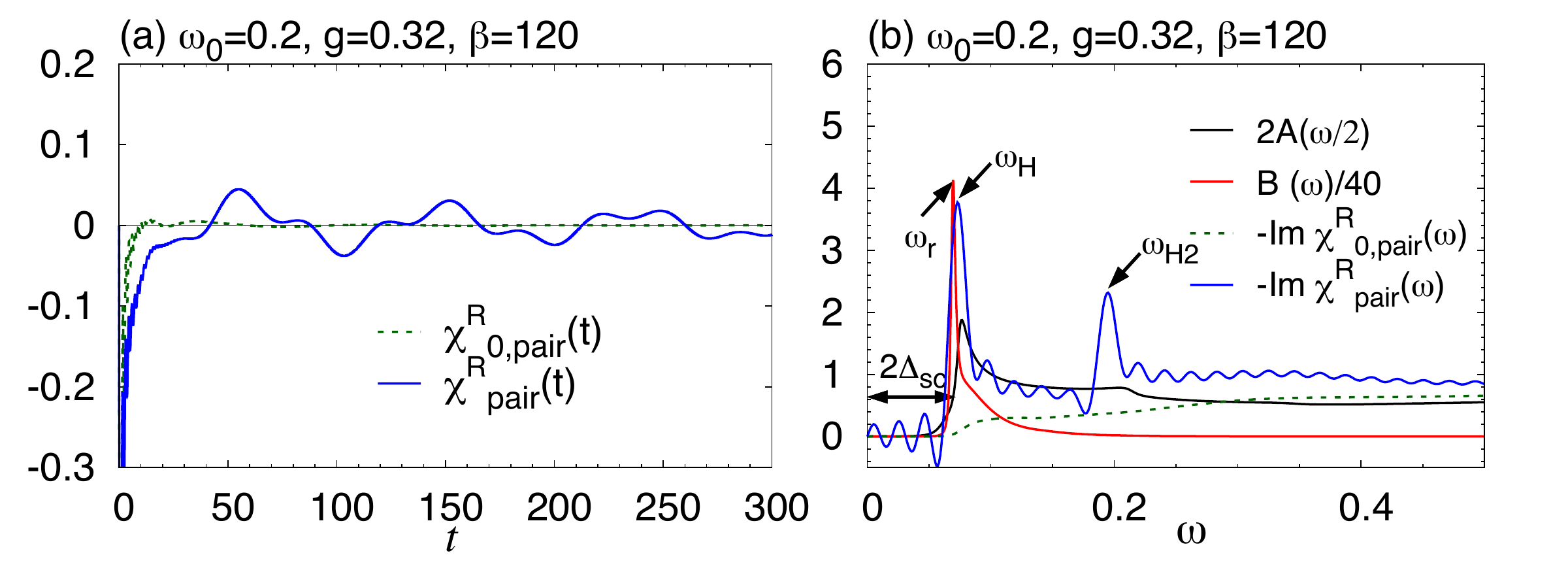}
  \caption{(a) Dynamical pair susceptibility against $t$ evaluated with the full dynamics in the self-consistent Migdal approximation [$\chi_{\rm pair}(t)$] and with 
the bubble diagrams [$\chi_{0,{\rm pair}}(t)$] for $\omega_0=0.2,g=0.32, \beta=120$ ($\lambda_{\rm eff}=1.65$). (b) Comparison of
the electron spectrum $A(\omega/2)$, the phonon spectrum $B(\omega)$, $-{\rm Im}\chi^R_{0,{\rm pair}}(\omega)$ and $-{\rm Im}\chi^R_{\rm pair}(\omega)$  
 for $\omega_0=0.2,g=0.32, \beta=120$. $\chi_{\rm pair}(\omega)$ and $\chi_{0,{\rm pair}}(\omega)$ are evaluated from the data at $t\in[0,300]$. }
  \label{fig:lam_eff_w0.2}
\end{figure}

In Fig.~\ref{fig:lam_eff_w0.2}, we show the result for $\omega_0=0.2$. The result involves totally similar features as for $\omega_0=0.4$, see Fig.~\ref{fig:suscep_pair}.
This fact indicates that the discussions made in the main part are applicable to lower phonon frequencies, where the Migdal approximation
 becomes quantitatively more reliable.
However, we note that systematic analyses for lower phonon frequencies than $\omega=0.4$ are difficult. 
This is because all dynamics involved becomes slower and hence numerical simulation becomes more demanding.

\section{Implementation of the pulse field}\label{sec:implement}
The Dyson equations involved in DMFT of the Holstein model with the pair potential, $H_{\rm ex}(t)=F_{\rm ex}(t) \sum_i(c_{i\uparrow}^\dagger c_{i\downarrow}^\dagger+c_{i\downarrow}c_{i\uparrow})$, are
\begin{align}
&D(t,t')=D_0(t,t')+[D_0*\Pi*D](t,t'),\label{eq:ph_dyson}\\
&\begin{bmatrix}
i\partial_{t}+\mu&- F_{\rm ex}(t)\\
-F_{\rm ex}(t)&i\partial_{t}-\mu
\end{bmatrix}
\hat{G}(t,t')
-[(\hat{\Sigma}+\hat{\Delta})* \hat{G}](t,t')=\hat{I}\delta_{\mathcal C} (t,t'),\\
&\begin{bmatrix}
i\partial_{t}+\mu&- F_{\rm ex}(t)\\
-F_{\rm ex}(t)&i\partial_{t}-\mu
\end{bmatrix}
\hat{\mathcal{G}}_0(t,t')
-[\hat{\Delta}* \hat{\mathcal{G}}_0](t,t')=\hat{I}\delta_{\mathcal C} (t,t').
\end{align}
Here $\hat{I}$ is the identity matrix, and $\hat{\Delta}(t,t')$ is the hybridization function, which is $v^2\hat{\sigma}_3\hat{G}(t,t')\hat{\sigma}_3$ on the Bethe lattice.
When we take $F_{\rm ex}(t)=d_{\rm f}\delta(t)$, one finds from the above Dyson equations that the effect of the external field leads to a jump in $\hat{\mathcal{G}}_0$ and $\hat{G}$ around $t=0$:
\begin{align}
& \hat{G}^R(0^+,0^+)=-i\hat{I},\label{eq:ret_jump}\\
&\hat{G}^\rceil(0^+,\tau')
=\hat{M}\hat{G}^\rceil(0^-,\tau'),\label{eq:leftmix_jump}\\
&\hat{G}^<(0^+, 0^+)
=\hat{M}\hat{G}^<(0^-,0^-)\hat{M}^\dagger,\label{eq:lesser_jump}
\end{align}
where we have defined the matrix $\hat M$,
\begin{align}
\hat{M}\equiv\frac{1}{1+\frac{d_f^2}{4}}
\begin{bmatrix}
1-\frac{d_f^2}{4}&-id_f\\
-id_f&1-\frac{d_f^2}{4}
\end{bmatrix}.
\end{align}
The expressions for the discontinuity of the Weiss Green's functions are 
obtained by replacing $G$ with $\mathcal{G}_0$ in Eq.~(\ref{eq:ret_jump}), (\ref{eq:leftmix_jump}), (\ref{eq:lesser_jump}).
On the other hand, the phonon Green's function ($D$) is continuous there.

\section{Dynamical pair susceptibility}\label{sec:chi}
The dynamical pair susceptibility can be expressed by the response of the Green's functions to modulations of the pair potential as in Eq.~(\ref{eq:pair}). 
Hence we want to calculate the quantity,
\begin{align}
\hat{\Lambda}_{\bf k}(t,t';t'')\equiv \left.\frac{\delta_{\mathcal C} [\hat{G}_{\bf k}(t,t')]}{{\delta_{\mathcal C} [F_{\rm ex}(t'')}]}\right|_{F_{\rm ex}(t)=0}.
\end{align}

In the case of a free system, this quantity becomes
\begin{align}
\hat{\Lambda}_{0,\bf k}(t,t';t'')=\hat{G}_{0,\bf k}(t,t'')\hat{\sigma}_1\hat{G}_{0,\bf k}(t'',t'),
\end{align}
where the suffix $0$ denotes bare propagators.
For general interacting cases, we introduce the vertex part $\hat{\Gamma_{\bf k}}$ as
\begin{align}
\hat{\Lambda}_{\bf k}(t,t';t'')=\int_{\mathcal C} d{t_1}d{t_2}\hat{G}_{\bf k}(t,t_1)\hat{\Gamma}_{\bf k}(t_1,t_2;t'')\hat{G}_{\bf k}(t_2,t')\label{eq:vertex_def}.
\end{align}
In the following, we assume that the self-energy is momentum independent (DMFT approximation). From the Dyson equation for $\hat{G}_{\bf k}$, it then follows that 
\begin{align}
&\hat{\Lambda}_{\bf k}(t,t';t'')=\hat{\Lambda}_{0,\bf k}(t,t';t'')\nonumber\\
&+\int_{\mathcal C}  d{t_1}d{t_2} \hat{\Lambda}_{0,\bf k}(t,t_1;t'')\hat{\Sigma}(t_1,t_2)\hat{G}_{\bf k}(t_2,t')\nonumber\\
&+\int_{\mathcal C} d{t_1}d{t_2} \hat{G}_{0,\bf k}(t,t_1)\left.\frac{\delta_{\mathcal C} [\hat{\Sigma}(t_1,t_2)]}{\delta_{\mathcal C} [F_{\rm ex}(t'')]}\right|_{F_{\rm ex}=0}\hat{G}_{\bf k}(t_2,t')\label{eq:lam_eq}\nonumber\\
&+\int_{\mathcal C} d{t_1}d{t_2} \hat{G}_{0,\bf k}(t,t_1)\hat{\Sigma}(t_1,t_2)\hat{\Lambda}_{\bf k}(t_2,t';t'').
\end{align}
From Eqs.~(\ref{eq:vertex_def}),(\ref{eq:lam_eq}) and the Dyson equation, we obtain the expression for the vertex part Eq.~(\ref{eq:vetex_general}).
One notices that the vertex does not depend on ${\bf k}$, either. 

For diagrammatic approximations we explicitly know the expression for the self-energy, hence we can directly determine the vertex correction, $\left.\frac{\delta_{\mathcal C} [\hat{\Sigma}(t,t')]}{\delta_{\mathcal C} [F_{\rm ex}(t'')]}\right|_{F_{\rm ex}=0}$ from it. 
In the present case of DMFT+self-consistent Migdal approximation, the self-energies for 
electrons ($\hat{\Sigma}$) and phonons ($\Pi$) are expressed as Eq.~\ref{eq:Migdal}.
From this we obtain
\begin{align}
\frac{\delta_{\mathcal C} [\hat{\Sigma}(t,t')]}{\delta_{\mathcal C} [F_{\rm ex}(t'')]}=&ig^2D(t,t')\hat{\Lambda}(t,t';t'')\nonumber\\
&+ig^2\Omega(t,t';t'')\hat{\sigma}_3\hat{G}(t,t')\hat{\sigma}_3,
\end{align}
where we have defined 
\begin{align}
\Omega(t,t';t'')&\equiv\frac{\delta_{\mathcal C} [D(t,t')]}{\delta_{\mathcal C} [F_{\rm ex}(t'')]}\nonumber\\
&\equiv\int_{\mathcal C} dt_1dt_2 D(t,t_1)\Theta(t_1,t_2;t'')D(t_2,t')\label{eq:Omega}
\end{align}
and, as in the main text, $\hat{\Lambda}(t,t';t'')\equiv\frac{1}{N}\sum_{\bf k}\int_{\mathcal C} d{t_1}d{t_2}\hat{\sigma_3}\hat{G}_{\bf k}(t,t_1)\hat{\Gamma}(t_1,t_2;t'')\hat{G}_{\bf k}(t_2,t')\hat{\sigma_3}$. 
From the Dyson equation for the phonon Green's function Eq.~(\ref{eq:ph_dyson}) and with the same procedures as for $\hat{\Lambda}_{\bf k}$, we find
\begin{align}
&\Theta(t,t';t'')=\frac{\delta_{\mathcal C} [\Pi(t,t')]}{\delta_{\mathcal C} [F_{\rm ex}(t'')]}\nonumber\\
&=-ig^2\{{\rm tr}[\hat{\Lambda}(t,t';t'')
\hat{G}(t',t)]+{\rm tr}[\hat{G}(t,t')\hat{\Lambda}(t',t;t'')]\}.
\end{align}
Hence the final expression for the vertex function becomes Eq.~(\ref{eq:vertex}) in the main part.
The derivation of Eq.~(\ref{eq:vetex_uMig}) for the unrenormalized Migdal approximation is similar but much simpler.

\end{document}